\documentclass[%
 reprint,
 amsmath,amssymb,
 aps,superscriptaddress,floatfix
prl,
]{revtex4-2}
\usepackage{color}
\usepackage{graphicx}
\usepackage{hyperref}
\usepackage{natmove}
\usepackage{dcolumn}
\usepackage{bm}

\begin{document}

\preprint{APS/123-QED}

\title{Occupational Disorder as the Origin of Flattening of the Acoustic Phonon Branches in the Clathrate Ba$_{8}$Ga$_{16}$Ge$_{30}$}

\author{Susmita Roy}
\affiliation{Department of Physics, University of Colorado Boulder, 390 UCB, Boulder, 80309, Colorado, USA}

\author{Tyler C. Sterling}
\affiliation{Department of Physics, University of Colorado Boulder, 390 UCB, Boulder, 80309, Colorado, USA}

\author{Daniel Parshall}
\affiliation{Department of Physics, University of Colorado Boulder, 390 UCB, Boulder, 80309, Colorado, USA}
\affiliation{Department of Economics, Universidad del Rosario, Calle 12C 6-25, Bogota, 111711, D.C., Colombia}

\author{Eric Toberer}
\affiliation{Department of Physics, Colorado School of Mines, 1500 Illinois St., Golden, 80401, Colorado, USA}

\author{Mogens Christensen}
\affiliation{Department of Chemistry and Interdisciplinary Nanoscience Center (iNANO), Aarhus University, Langelandsgade 140, Aarhus,
8000, Denmark}

\author{Devashibhai T. Adroja}
\affiliation{ISIS Facility, STF, Rutherford Appleton Laboratory, , Chilton, 610101, Oxfordshire OX11 0QX, United Kingdom}

\author{Dmitry Reznik}
\email{dmitry.reznik@colorado.edu}
\affiliation{Department of Physics, University of Colorado Boulder, 390 UCB, Boulder, 80309, Colorado, USA}
\affiliation{Center for Experiments on Quantum Materials, University of Colorado Boulder, 390 UCB, Boulder, 80309, Colorado, USA}

\date{\today}

\begin{abstract}
In the search for high-performance thermoelectrics, materials such as clathrates have drawn attention due to having both glass-like low phonon thermal conductivity and crystal-like high electrical conductivity. Ba$_{8}$Ga$_{16}$Ge$_{30}$ (BGG) has a loosely bound guest Ba atom trapped inside rigid Ga/Ge cage structures. Avoided crossings between acoustic phonons and the flat guest atom branches have been proposed to be the source of the low lattice thermal conductivity of BGG. Ga/Ge site disorder with Ga and Ge exchanging places in different unit cells has also been reported. We used time-of-flight neutron scattering to measure the complete phonon spectrum in a large single crystal of BGG and compared these results with predictions of density functional theory to elucidate the effect of the disorder on heat-carrying phonons. Experimental results agreed much better with the calculation assuming the disorder than with the calculation assuming the ordered configuration. Although atomic masses of Ga and Ge are nearly identical, we found that disorder strongly reduces phonon group velocities, which significantly reduces thermal conductivity. Our work points at a new path towards optimizing thermoelectrics. 
\end{abstract}

\maketitle



Thermoelectric materials enable environmentally-friendly waste-heat to electricity conversion \cite{snyder2008complex, zhang2015thermoelectric}. The efficiency of a thermoelectric is determined by a dimensionless quantity called the figure of merit: $ZT=\sigma S^{2} T/ \kappa$, in which $\sigma$ is the electrical conductivity, $S$ is the Seebeck coefficient, $T$ is the temperature, and $\kappa$ is the thermal conductivity \cite{ziman1972principles}. The thermal conductivity $\kappa=\kappa_L+\kappa_e$ is the sum of lattice thermal conductivity ($\kappa_{L}$) and electronic thermal conductivity ($\kappa_{e}$). It is classically known that the lattice component is the relevant one for thermoelectric performance \cite{chasmar1959thermoelectric}. According to the kinetic theory of gases, $\kappa_{L}$ is proportional to the lattice heat capacity $C$, average phonon velocity $v$, and the average relaxation time $\tau$ of phonons. 
In most cases low, glass-like $\kappa_{L}$ is attributed to reduced $v$ and/or $\tau$.

\begin{figure*}
\includegraphics[width=1\linewidth]{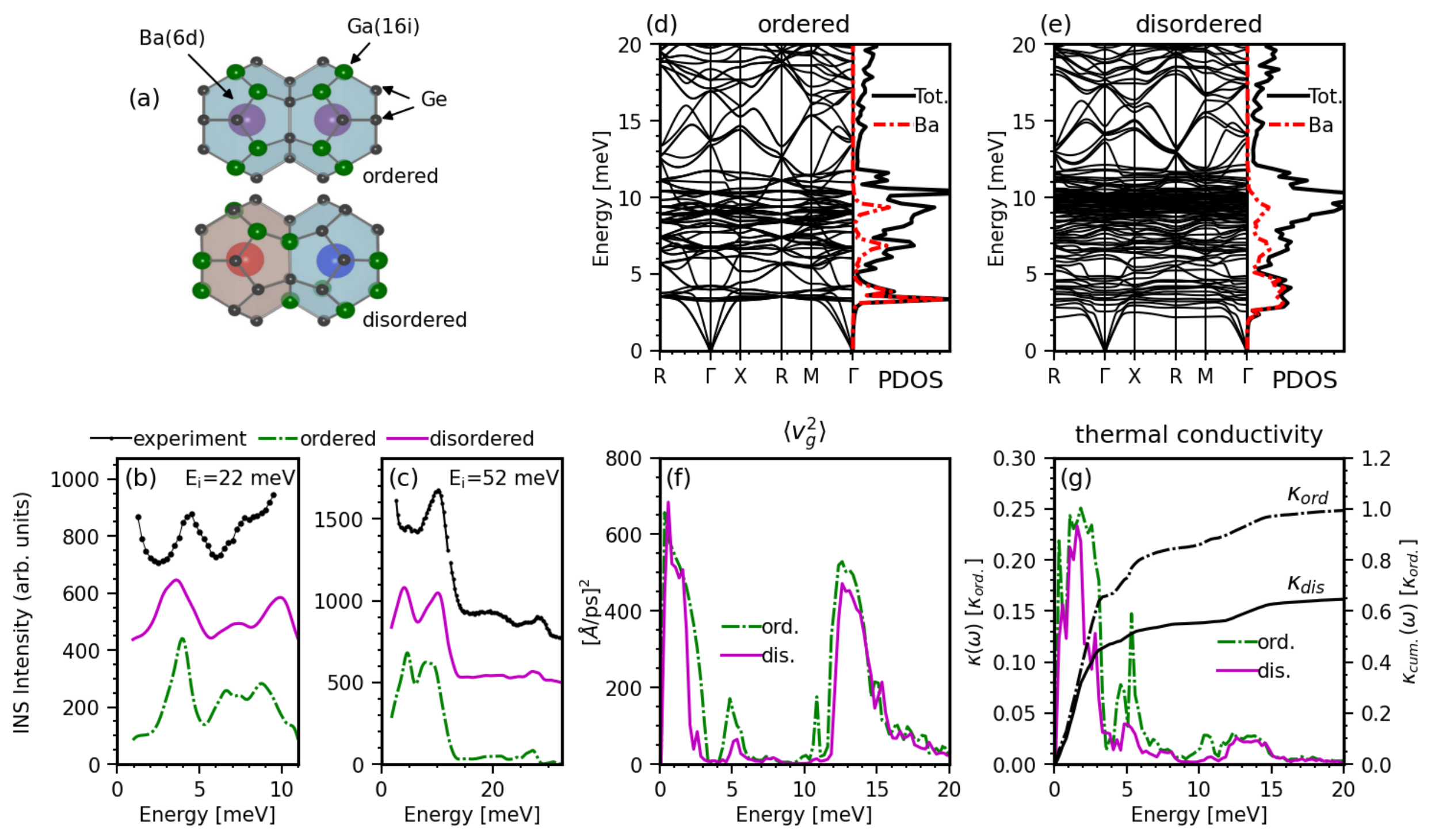}
    \caption{Summary of the lattice dynamical phenomena considered in this work. (a) Ba-6d containing (Ga, Ge)$_{24}$ cage portion of the 3d unitcells of perfectly ordered (top) and disordered, with scrambled Ge/Ga atom occupations (bottom) Ba$_8$Ga$_{16}$Ge$_{30}$.(b,c) Black lines show neutron-weighted densities of states (DoS) (inelastic neutron scattering intensities integrated over the 3$<$H$<$6, 3$<$K$<$7, 4.5$<$L$<$6.5 r.l.u. region of reciprocal space) measured with incident energy E$_i$=22 meV (b) and E$_i$=52 meV (c). The green/pink lines are the same quantity calculated from DFT using the ordered/disordered unit cells and broadened with an approximate resolution function fit to the experimental intensity (see supplementary information \cite{supp_info}). The disordered calculation and the experimental curves for E$_i$=22 meV are offset vertically by 350 and 400 counts respectively. The experimental curve has an additional large ($\sim$250 counts) background that is not present in the calculations. For E$_i$=52 meV, the disordered and experimental curves are offset by 500 and 700 counts respectively. Here, the experimental background is about $\sim$ 600 counts. (d,e) The ordered (d) and disordered (e) phonon dispersions and densities of states (PDOS). The dash-dot red curve is the projection onto the Ba atoms; the black curve is the total DOS. (f) The average group velocities squared, $\langle v^2_g\rangle$, calculated from the dispersions in (d) and (e) by averaging over all modes and $\bm{q}$-points. (g) Black curves represent the spectral, $\kappa(\omega)$, and colored curves represent the cumulative, $\kappa_{cum.}(\omega)$, thermal conductivities at 300 K calculated from the group velocities and densities of states in (f) and (d),(e). Only Umklapp processes are considered; $\tau(\omega)=\tau_0 \omega^{-2}$ with $\tau_0$ the same for the ordered and disordered calculations. The data in (g) are in units of the total thermal conductivity of the ordered crystal.}
    \label{fig:summary_fig}
\end{figure*}

Glass-like low lattice thermal conductivity and crystal-like high electrical conductivity coexist in so-called phonon-glass electron-crystal (PGEC) materials \cite{slack1995crc}. One way to realize this concept is by designing materials such that loosely bound ``guest'' atoms sit inside empty spaces of a rigid atomic lattice with good electronic conductivity \cite{nolas1998semiconducting,keppens1998localized,hsu2004cubic,poudeu2006high}. In this case interactions between the low-energy rattling motion of the guest atoms and acoustic phonons reduce $v$ and/or $\tau$ depending on specific materials. In particular, the reduction of $v$ is achieved through avoided crossings (anticrossings) between low-lying optic branches with acoustic branches that result from coupling between the phonons in branches that would cross in the absence of such a coupling. 
In this case the dispersion curves never cross instead "avoiding" each other. As a result the branches become more flat and $v$ decreases. 

The X$_8$Ga$_{16}$Ge$_{30}$ (X=Ba,Eu,Sr) clathrates have low lattice thermal conductivity ($\sim$ 1 W/mK at room temperature), which makes them promising candidates for efficient thermoelectrics \cite{nolas1998semiconducting,bentien2004thermal,cohn1999glasslike}. Their structure is characterized by tetrakaidecahedral (Ga, Ge)$_{24}$ and dodecahedral cages (Ga, Ge)$_{20}$ with loosely bonded Ba, Eu, or Sr guest atoms inside, which makes them a classic type of PGEC materials \cite{kovnir2004semiconducting,toberer2010zintl}. Evidence of strong occupational disorder of the Ga/Ge sites has also emerged recently, i.e. Ga and Ge are distributed nearly randomly on the cage vertices [fig. \ref{fig:summary_fig} (a)]  \cite{chakoumakos2001structural,christensen2006crystal,bentien2000experimental,bentien2005crystal,bentien2002maximum}. Since Ga and Ge have similar atomic masses, the effect of their occupational disorder on phonon dispersions was implicitly assumed to be minimal. 
Here we combined comprehensive time-of-flight (TOF) inelastic neutron scattering measurements of Ba$_{8}$Ga$_{16}$Ge$_{30}$ with calculations based on the density functional theory (DFT) assuming both the ordered and the disordered structures and found that the disorder splits the degenerate rattler branches into multiple nearly flat branches. The new branches produce many more avoided crossings with the acoustic modes, which significantly lowers their average group velocities and, as a consequence, reduces thermal conductivity. Our result demonstrates that occupational disorder control represents a new direction in the design of thermoelectric materials based on clathrates.



\begin{figure}
\includegraphics[width=1\linewidth]{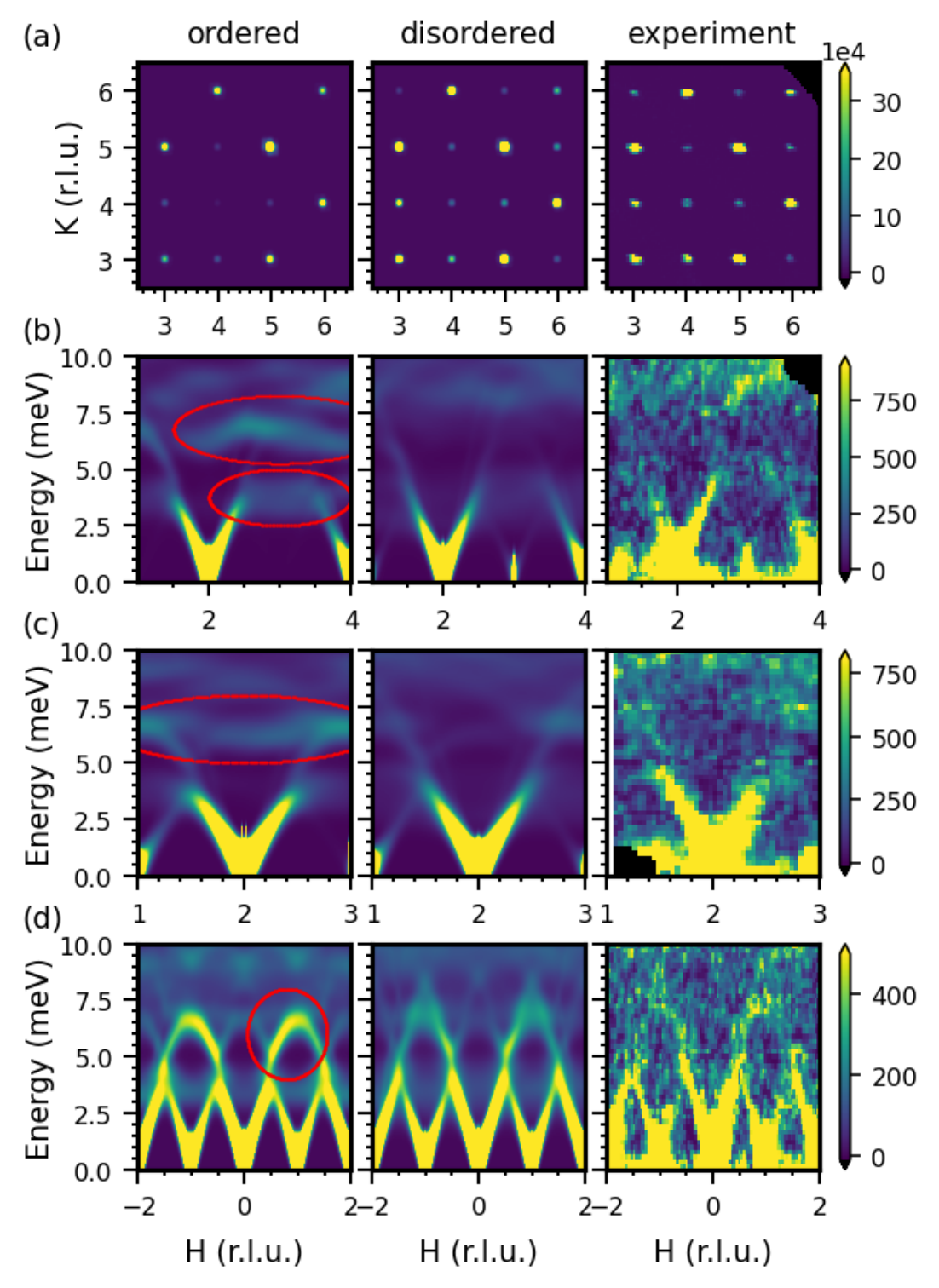}
    \caption{Ordered (left column) and disordered (middle column) neutron spectra compared to experiment (right column). (a) Bragg peaks (-1$<$E$<$1 meV) in the (H,K,L=6) plane. In the ordered cell, coherent scattering from the ordered arrangement of atoms results in certain Bragg peaks (e.g. $\textbf{Q}$=(4,4,6)) having no intensity. On the other hand, scattering from the disordered arrangement of atoms results in some remaining intensity at these Bragg peaks, in excellent agreement with the experimental Bragg pattern. The remaining rows show inelastic scattering spectra S($\textbf{Q},\omega$) along the $\textbf{Q}$=(H,6,6) (b), $\textbf{Q}$=(H,2,8) (c), and $\textbf{Q}$=(H,5,3) (d) reciprocal lattice directions. The disordered calculation is averaged over all otherwise equivalent directions as explained in the text. Red ovals in the ordered calculation indicate intensity from excitations that is not visible in the disordered calculation and experiment as discussed below.}
    \label{fig:bragg_sqw}
\end{figure}

\begin{figure}
\includegraphics[width=1\linewidth]{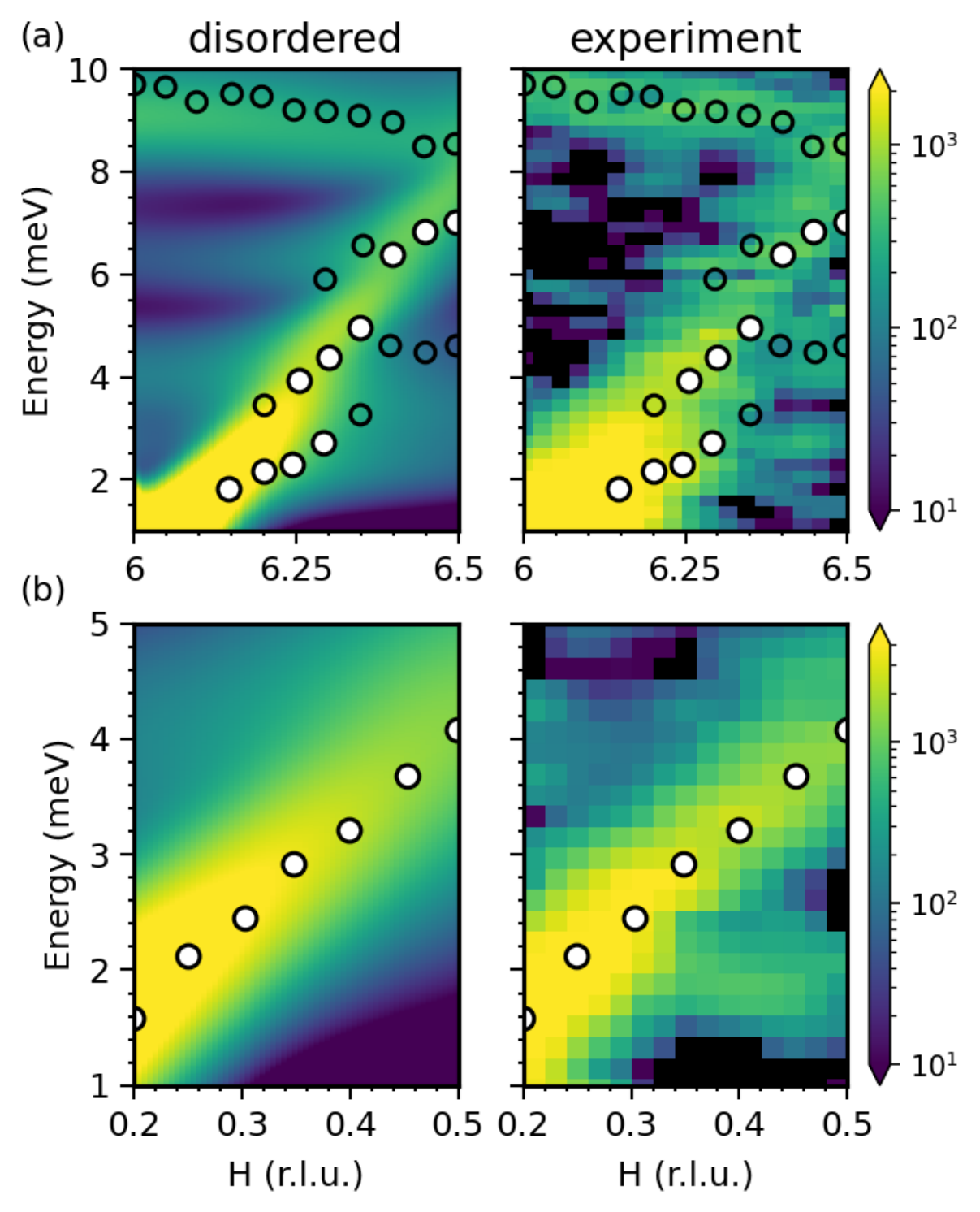}
    \caption{(a) LA and (b) TA phonon scattering intensities (color maps) at $\textbf{Q}$=(5.5+h,0,0) (a) and $\textbf{Q}$=(6,h,0) (b) and experimental phonon energies (circles) obtained from the multizone fit of experimental inelastic neutron scattering intensity S(Q,$\omega$) (see fig. $5$ in supplementary information). The big circles indicate acoustic phonons and the small circles are optic modes. We do not include the small intensity near 2.5 meV in our phonon fits as it does not appear in any other zones and may be an artifact. }
    \label{fig:dispersions}
\end{figure}

Neutron scattering measurements were performed on the MERLIN direct geometry chopper spectrometer at the ISIS Neutron and Muon source at the Rutherford Appleton Laboratory in Didcot, UK \cite{bewleyMERLINNewHigh2006}. The Ba$_{8}$Ga$_{16}$Ge$_{30}$ crystal used for the inelastic neutron scattering experiment is the same sample that was used in ref. \cite{christensen2008avoided}. MERLIN has high flux and a large detector area, collecting the four dimensional inelastic neutron scattering data set S($\bf{Q}$,$\omega$) across many Brillouin zones (BZ). For data analysis, we used the Phonon Explorer software \cite{phonon-explorer,reznik2020automating}, which enables efficient search for wavevectors where a particular effect is observed most clearly, and performs multizone fitting to efficiently separate phonon branches that are much more closely spaced than the experimental resolution \cite{parshall2014phonon}. We used \textsc{phonopy} to solve the lattice dynamical equations based on the force-constants from Ref. \cite{ikeda2019kondo}. The inelastic neutron scattering structure factors, S($\bf{Q}$,$\omega$) were simulated using the  \textsc{snaxs} \cite{parshall} and \textsc{euphonic} \cite{euphonic} softwares. To represent the finite line widths and experimental resolution broadening, the structure factors in figs. \ref{fig:bragg_sqw} and \ref{fig:dispersions} were convoluted with a Gaussian function assuming the full-width at half-maximum (FWHM) is 0.75 meV. The intensity scales in arbitrary units for all S($\bf{Q}$,$\omega$) calculations are the same.

The ordered phase of Ba$_{8}$Ga$_{16}$Ge$_{30}$ typically used in DFT calculations (fig. \ref{fig:summary_fig} (a), top) has cubic symmetry and intensities along reciprocal lattice directions with permuted axes are identical. On the other hand, the real structure of Ba$_{8}$Ga$_{16}$Ge$_{30}$ is disordered with Ga and Ge atoms randomly distributed on the cage vertices (fig. \ref{fig:summary_fig} (a), bottom) \cite{chakoumakos2001structural,christensen2006crystal,bentien2000experimental,ikeda2019kondo,bentien2005crystal,bentien2002maximum}. The disorder breaks the cubic symmetry (Pm-3n$\rightarrow$P1) and intensities along directions that are equivalent in the ordered crystal are no longer identical. Still, the experimental S($\bf{Q}$,$\omega$) is averaged over the large ($\sim 10^{23}$) number of different disordered unit cell configurations of the macroscopic crystal, which results in an apparent cubic symmetry.

In DFT the force constants of Ba$_{8}$Ga$_{16}$Ge$_{30}$ were calculated from a single unit cell in both the ordered and disordered phases \cite{ikeda2019kondo} because Ba$_{8}$Ga$_{16}$Ge$_{30}$ has a large enough unit cell that force-constants fall off sufficiently at the cell boundary. Still, disorder breaks translational symmetry, so the computational unit cell should be sufficiently large that atoms on opposite sides of the cell are uncorrelated. Moreover, one should calculate phonons using an ensemble of disordered configurations and average over all ensembles. However, this is too computationally expensive in our case. 

We calculated S($\bf{Q}$,$\omega$) from a single configuration (from ref. \cite{ikeda2019kondo}) chosen using the ``special quasi-random structure" (SQS) method \cite{zunger1990special}. The goal of SQS is to pick a small computational supercell that best matches the disorder in a very large (ideally infinite) supercell. This was achieved by fitting correlation functions calculated from a single unit cell with scrambled Ge/Ga occupations to true ``random" correlation functions; e.g. from experiment or calculated from a very large supercell. The ``disorder" (in the case of Ba$_8$Ga$_{16}$Ge$_{30}$, the Ga/Ge site occupancies) is varied to minimize the difference between the correlation function(s) of the computational cell and the true random correlation function(s). The structure with the best match was chosen for the calculations. To approximate the implicit directional averaging, the theoretical S($\bf{Q}$,$\omega$) calculated from this disordered cell were averaged over all directions that are equivalent assuming cubic symmetry.

Calculation based on the ordered structure where the optic flat branches bunch into several narrow energy intervals. Projecting the phonon densities of states onto the individual atoms shows primarily Ba - 6d, character, thus these are the rattler modes of Ba in the 6d site. Disorder spreads these branches out in energy (Fig. \ref{fig:summary_fig}d,e). In particular, the ordered calculation gives avoided crossings between acoustic branches and the nearly flat rattler branches near $\sim$4 meV, which tend to reduce the phonon group velocity and, as a consequence, the thermal conductivity. Due to disorder, these avoided crossings become distributed throughout a broad energy interval between 2 and 5 meV, which futher suppresses thermal conductivity by a large amount. The main effect of the splitting of the rattler modes is to create additional avoided crossings with the acoustic phonons (see supplementary information for a more detailed discussion \cite{supp_info}).

Fig. \ref{fig:summary_fig}b-c shows that the calculated phonon density of states (DOS) corrected for the neutron cross section agrees much better with experiment when the disordered structure is used. In both calculations the energy of the lowest DOS peak is about 3meV vs 4.5 meV in experiment, consistent with the known tendency of the DFT to underestimate the force constants overall. However, the width of this peak, which comes from the gaps at the avoided crossings of acoustic phonons and rattler modes, is increased in the disordered calculation to closely match experiment. Moreover, the calculated Bragg peak intensities agree much better with experiment if disorder is included in the calculation (Fig. \ref{fig:bragg_sqw}). The same applies to select phonon spectra (Fig. \ref{fig:bragg_sqw}b-d). Color maps of the simulated and background-subtracted experimental spectra of acoustic phonons shown in Figure \ref{fig:dispersions} are similar, highlighting the accuracy of the calculation. Peak positions obtained from the multizone fit (see section V in supplementary information for details) agree very well with the color maps of both the simulation and the experiment. In particular, signatures of avoided crossings of the LA branch with optic modes are clearly visible in the simulation and the data.

Our calculation of thermal conductivity (Fig. \ref{fig:summary_fig}f-g) considered Umklapp scattering only. A more accurate calculation just for the disordered case is reported in \cite{ikeda2019kondo}. Fig. \ref{fig:summary_fig}f-g illustrates a profound effect of the disorder, which suppresses phonon thermal conductivity mostly around 3-5meV where the effect of the rattler branches is the strongest.

Comparison of our calculations to the Raman results of Ref. \cite{takasu2006dynamical} shows that broadening of the Raman phonon peaks far beyond the experimental resolution is a natural consequence of disorder-induced splitting of the modes \cite{supp_info}.


In Ba$_8$Ga$_{16}$Ge$_{30}$ and similar semiconducting clathrates, most DFT calculations investigating the avoided crossing regions assume that the Ga/Ge site occupancies are fully ordered with Ga only in the 16i site and Ge in the remaining cage vertex positions \cite{tadano2015impact,euchner2019predicting,gonzalez2017estimating,tadano2018quartic,blake1999clathrates,dong2000chemical}. However, there is extremely strong evidence that the Ga/Ge occupancies on the cages vertices are disordered \cite{chakoumakos2001structural,christensen2006crystal,bentien2000experimental,ikeda2019kondo,bentien2005crystal,bentien2002maximum}. 
Unfortunately modeling disordered materials from first principles greatly increases the computational workload, which limited the number of studies of the lattice dynamics in clathrates \cite{he2014nanostructured,gao2017giant}.

Ikeda et al. \cite{ikeda2019kondo} investigated the \emph{temperature-dependence} of the thermal conductivity and specific heat of Ba$_8$Ga$_{16}$Ge$_{30}$ with and without disorder. Only when accounting for correlation in a Kondo-like phonon effect could the experimental temperature dependence be explained \cite{ikeda2019kondo}. They noted that whereas the temperature dependence is robust against disorder, the absolute value of the thermal conductivity is lowered in the disordered calculation. The decrease was attributed to shorter lifetimes due to increased anharmonicity \cite{ikeda2019kondo}, similar to Ba$_8$Ga$_{41}$Au$_{5}$ where disorder increases the available phase-space for 3-phonon scattering,  \cite{lory2017direct}. Here we showed that another important effect of disorder is to lower the phonon group velocities.

Disorder-induced dispersion flattening was theoretically predicted in K$_8$Al$_8$Si$_{38}$, with Al/Si occupational disorder \cite{he2014nanostructured}. Another study found that isotope disorder in the guest atom sites in Si$_x$ and Ba$_8$Si$_x$ ($x\in\{46,230,644\}$) played a major role in reducing thermal conductivity \cite{gao2017giant} due to both lifetimes and group velocity reduction.  However, these effects have not been confirmed in experiment, which we did here for Ba$_8$Ga$_{16}$Ge$_{30}$.

Our ordered cell calculations show that the flat branches in figs. \ref{fig:bragg_sqw} and \ref{fig:dispersions} near $\sim$4 and $\sim$6 meV come mainly from pure Ba guest atom modes (See Fig. \ref{fig:summary_fig} and the atom projected INS intensities in the supplementary information \cite{supp_info}). These multiply degenerate branches all contribute some intensity to the spectrum. Their intensities appear as a weak but visible flat branch across the BZ (red ovals in fig. \ref{fig:bragg_sqw}). It also shows up as a narrow, pronounced peak in the ordered cell neutron-weighted DoS in fig. \ref{fig:summary_fig}. However, absence of this intensity in the experiment above the background and the significantly improved agreement of the Bragg patterns of the disordered cell over the ordered cell, necessitates the disordered calculation.

In the disordered cell calculations, broken symmetry splits the Ba rattler atom branches. Since the wave vectors $\bf{Q}$ with permuted axes are no longer equivalent, S($\bf{Q}$,$\omega$) is averaged over all equivalent cubic directions in the simulation suppressing and broadening the simulated intensity of the flat modes. As a result, the only substantial intensity is near the acoustic branches, consistent with experiment. The sharp, pronounced peaks near $\sim$4 and $\sim$6 meV in the neutron-weighted DoS (fig. \ref{fig:summary_fig}) become broad and flat when calculated from the disordered cell. Notably, the neutron-weighted DoS curve calculated from the disordered cell agrees with the experimental data much better than the simulation without disorder.

In Ba$_{8}$Ga$_{16}$Ge$_{30}$, an avoided crossing between an acoustic phonon mode and the flat guest modes was theoretically predicted and it was claimed that the Ba guest atoms behave as local resonating scattering centers for the acoustic phonon modes in the region of the avoided crossing \cite{dong2001theoretical}. Subsequent triple axis neutron scattering measurements along the [h h 0] direction \cite{christensen2008avoided} found that the Ba rattler atoms flatten the phonon dispersions, reducing the group velocity $v$, rather than acting as a local resonant scattering center and reducing $\tau$.
In our data focusing on the [h 0 0] direction large-gap avoided crossings with different well separated rattler atom branches are apparent in the LA phonon dispersion (fig. \ref{fig:dispersions}a).
Meanwhile, the intensity in fig. \ref{fig:dispersions}b depicts a TA mode that is linear and continuously dispersing in both the experiment and in the simulation based on disordered calculation. However, our DFT calculations predict many 'small-gap' avoided crossings near $\sim$4 and $\sim$6 meV, that are washed out by the instrument resolution both in the experiment and in the simulation for both the TA branch and, away from the large-gap avoided crossings, the LA branch.  

Ga and Ge differ in nuclear-charge and mass by only $\sim3.5\%$, so the perturbation to the force-constants induced by mass or nuclear-charge disorder is insignificant. However, the bonds formed with Ga and Ge have different valence configurations. The electronic structure, which affects the Born-Oppenheimer electronic-energy part of the interatomic force constants, is thus qualitatively different between the ordered and disordered configurations.  The impact of disorder can be seen in the substantially lowered average group velocities due to small-gap avoided crossings, between $\sim 2$ and $\sim 4$ meV in fig. \ref{fig:summary_fig}). (Also see supplementary information for a heuristic explanation of this concept \cite{supp_info}). Regardless of any reductions in $\tau$ (which are probably also present \cite{ikeda2019kondo}), the velocity $v$ in the kinetic theory model is substantially reduced, which might explain the anomalously low thermal conductivity in Ba$_{8}$Ga$_{16}$Ge$_{30}$. 


To conclude, our calculations using the disordered unit cell validated by experiments show that Ga/Ge occupational disorder has a large effect on phonon dispersions through strong influence on electronic structure that underlies interatomic force constants. The disorder introduces many closely spaced optic branches with many more avoid crossings with acoustic modes. These small gaps are essential for understanding the observed low thermal conductivity of Ba$_{8}$Ga$_{16}$Ge$_{30}$ and other similar materials. 

All work at the University of Colorado was supported by U.S. Department of Energy, Office of Basic Energy Sciences, Office of Science, under Contract No. DE-SC0006939. EST acknowledges the support from NSF DMR 1555340. We thank the ISIS Facility for beam time RB1410509. We would like the thank Holger Euchner and Silke Paschen for providing us with the DFT force-constants from ref. \cite{ikeda2019kondo}.



\bibliography{thermo}

\end{document}